# Stochastic simulation of residential building occupant-driven energy use in a bottom-up model of the U.S. housing stock

Jianli Chen[1], Rajendra Adhikari[2], Eric Wilson[2], Joseph Robertson[2], Anthony Fontanini[2], Ben Polly[2], Opeoluwa Olawale[2]

[1] University of Utah
[2] National Renewable Energy Laboratory

## Abstract

The residential buildings sector is one of the largest electricity consumers worldwide and contributes disproportionally to peak electricity demand in many regions. Strongly driven by occupant activities at home, household energy consumption is stochastic and heterogeneous in nature. However, most residential building energy models applied by industry use homogeneous, deterministic occupant activity schedules, which work well for predictions of annual energy consumption, but can result in unrealistic hourly or sub-hourly electric load profiles, with exaggerated or muted peaks. This mattered less in the past, but the increasing proportion of variable renewable energy generators in power systems means that representing the heterogeneity and stochasticity of occupant behavior is crucial for reliable energy planning. This is particularly true for systems that include distributed energy resources, such as grid-interactive efficient buildings, solar photovoltaics, and battery storage. This work presents a stochastic occupant behavior simulator that models the energy use behavior of individual household members. It also presents an integration with a building stock model to simulate residential building loads more accurately at community, city, state, and national scales. More specifically, we first employ clustering techniques to identify distinct patterns of occupant behavior. Then, we combine time-inhomogeneous Markov chain simulations with probabilistic sampling of event durations to realistically simulate occupant behaviors. This stochastic simulator is integrated with ResStock™, a large-scale residential building stock simulation tool, to demonstrate the capability of stochastic residential building load modeling at scale. The simulation results were validated against both American Time Use Survey data and measured end-use electricity data for accuracy and reliability.

## 1. Introduction
### 1.1 Context and Motivation

The residential building sector constitutes one of the largest consumers of energy across the world, ranging from 15 to 28 percent according to the building energy consumption information review [1]. With urbanization, population growth, and increasingly digital and connected lifestyles, global energy consumption from the residential sector is expected to increase 1.4% annually over the next 30 years [2]. In the United States, the residential sector uses 38% of electricity [3], [4] and contributes disproportionally to peak electricity demand, representing over

50% of peak electricity demand in some regions of the country, such as Texas [5]. With the increasing supply of variable renewable energy into electricity systems, the timing and flexibility of electricity use is becoming more important; in some locations, wholesale electricity prices are going negative with increasing frequency [6], [7], and "duck curve" challenges are becoming more critical for utilities [8].

Residential building load simulation has typically been based on deterministic household activity schedules, which neglect the heterogeneity and stochasticity of resident behaviors. This approach has been sufficient in most cases for predictions of annual energy consumption, though for some end-use technologies, even the prediction of annual energy consumption requires a more stochastic and time-resolved modeling approach. Figure 1 (left) shows an example of a deterministic schedule for hot water usage, representing a blended average across many households, compared to an example of a stochastically generated hot water usage schedule for a single day in a single household (right). Wilson et al. [9] documents deterministic schedules covering all major end uses, which have been established as standard occupancy simulation protocols for the U.S. Department of Energy's Building America program.

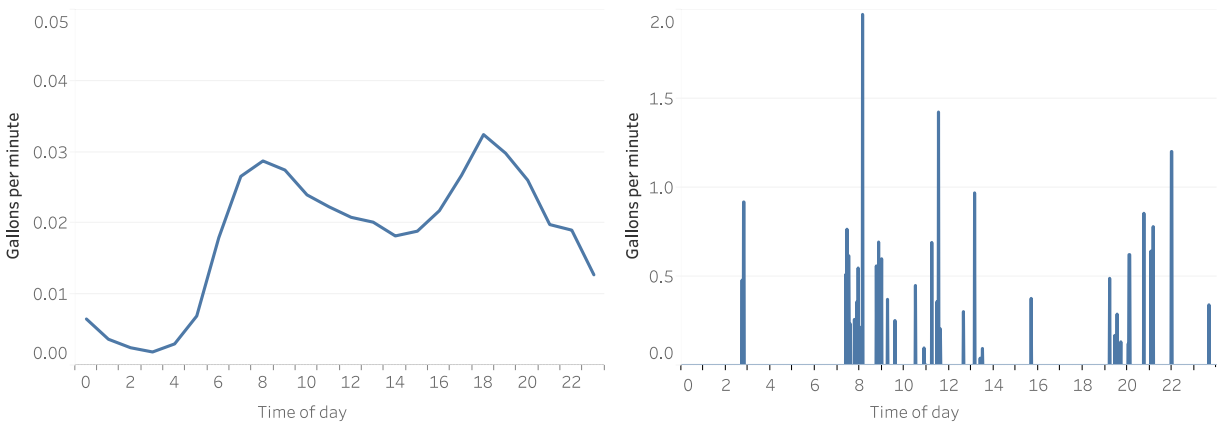

*Figure 1. Example of a deterministic schedule for daily hot water usage, derived from an average across many households (left) and a stochastically generated hot water usage schedule for a single day in a single household (right). Note that the vertical axis scale on the right-hand plot is 40 times that of the left-hand plot.*

Traditionally, residential building energy simulation has focused on predicting *annual* energy consumption (e.g., for rating and labeling systems) and on predicting *annual* energy savings from energy efficiency measures (e.g., for energy efficiency program design). Using deterministic and homogeneous behavior schedules is generally well-suited for these applications, with some exceptions for certain technologies, such as heat pump water heaters [10]. Even when *hourly* simulation results are of interest, using a fixed schedule that represents an average of all households is appealing, because it allows the use of a single model (or a handful of prototype models) to represent the entire residential sector, with fairly accurate estimates of sector-wide hourly electricity demand by end use. Such an approach was used to produce a dataset that has been widely used by industry and academia since 2013 [11].[1]

However, this homogeneous, deterministic approach to representing occupant behavior fails to capture the diversity of how individual households are impacted by various technologies and policies. Frick et al. [12] consulted with a group of more than 70 individuals representing more than 50 organizations in the United States, including utilities, independent system operators/regional transmission organizations, public utility commissions, state and local government, consulting firms, software companies, academic institutions, nongovernmental organizations that represent utilities and regional efficiency groups, national laboratories, and the U.S. Department of Energy. They found that a majority of use cases (35 out of 61) for end-use load profile modeling either require, or would benefit from, individual building or customer end-use load profiles exhibiting realistic degrees of heterogeneity and stochasticity [12]. Examples of these use cases include utility rate design and ratepayer impact analyses, electric distribution system analyses (non-wires alternatives, electric vehicle and solar photovoltaic hosting capacity), solar photovoltaic cost effectiveness analyses, behind-the-meter battery and inverter sizing, microgrid design, and demand response and demand flexibility potential studies.

### 1.2 Stochasticity and Heterogeneity

In this paper, we use *heterogeneity* to refer to the diversity of resident behavior and resulting electricity use, in terms of both magnitude and schedule. Heterogeneity in behavior and energy use between households is driven by several factors, including differing numbers of occupants, occupants' employment or school schedules, occupants' preferences (e.g., cooking meals vs. dining out), the presence of appliances (e.g., dishwasher, in-unit laundry), thermal comfort preferences, and so on. Physical building characteristics (e.g., insulation levels), weather, and geographic location (e.g., sunrise/sunset timing) contribute to heterogeneity in the magnitude and timing of energy use between households—particularly for heating, air conditioning, and lighting energy use. However, diversity in these parameters is already well-characterized in our modeling framework and thus is not a focus of this paper [13].

We use the term *stochasticity* to represent the potential variance of occupant behaviors over time (i.e., from day to day and week to week). A stochastic model can be used to achieve both heterogeneity across households and variability over time [14]. However, using a stochastically generated profile applied to multiple households without sufficient heterogeneity will likely lead to coincident events that exaggerate the peak load and generate unrealistic aggregate load shapes, causing inaccurate analysis results for electrical utilities, cities, and grid operators. Hence, it is important to account for both heterogeneity and stochasticity to realistically simulate residential building loads in different applications.

### 1.3 Literature Review

Traditionally, both top-down and bottom-up approaches have been used to estimate residential building loads. Langevin et al. provide a comprehensive classification for different types top-down and bottom-up building stock models [15]. The top-down approaches model residential building energy consumption from a holistic point of view and associate residential energy use with economic factors (e.g., gross domestic product or household income), human factors (household member composition), and physical factors (e.g., appliance saturation rates) [16]. For example, Aigner et al. [17] developed an hourly load estimate model for a group of residential buildings using regression techniques, such as conditional demand analysis (CDA), by accounting for a variety of variables, such as the outdoor air temperature, dwelling size, and household appliance characteristics. Similarly, Bartels et al. [18] simulated hourly building load profiles using CDA, with weather and socioeconomic variables as inputs, and then quantified the influence of climate change on building loads. Overall, considering aggregate characteristics, top-down models are suitable for providing a high-level analysis of residential building loads. Typically, only aggregate energy consumption data with easily accessible inputs are required in the modeling process. However, top-down models are not capable of providing detailed and individual end-use energy consumption profiles of residential buildings, which could hinder the application of these models in certain scenarios, such as retrofit analysis, which requires the evaluation of new technologies [19]. On the other hand, bottom-up approaches simulate building load profiles using input data for individual representative buildings or appliances. Due to their ability to explicitly provide details of disaggregated building loads and to consider diverse occupant behavior, bottom-up approaches, especially those using engineering-based models, have attained popularity among researchers and industrial practitioners [20].

---

[1] Based on 368 Google Scholar citations and personal correspondence between one of our authors and dozens of dataset users in industry and academia.

Engineering-based bottom-up approaches typically employ heat transfer calculations and end-use appliance modeling for building load simulation. Stochastic modeling of individual end-use loads has been developed to increase the fidelity of simulated residential building load profiles. For example, the highly cited work of Richardson et al. [21] established a high-resolution model to simulate domestic electricity usage of daily household activities (laundry, house cleaning, entertainment) and validated it using recorded household energy consumption in the UK. Similarly, Muratori et al. [22] developed a highly resolved modeling technique to simulate heating, ventilation, and air conditioning (HVAC) usage, lighting, and other household activities, which they then used to simulate overall residential electricity demand. Diao et al. [23] simulated residential energy consumption by incorporating major aspects of end-use consumption (HVAC, appliance usage, lighting) based on the American Time Use Survey (ATUS). These simulation results were later compared to simulations using ASHRAE end-use schedules and were found to achieve improved accuracy. Widen and Wäckelgård [24] tuned Markov chains to stochastically simulate household energy consumption using Swedish data. Ortiz et al. [25] developed a probabilistic approach to simulate household electricity in the Mediterranean region. Finally, Fisher et al. [26] derived probability distributions for household events and considered seasonal differences to simulate the electric load of residential buildings in Germany.

Generally, bottom-up stochastic residential building loads are simulated using either probability sampling or Markov chains [27]. The probability-sampling-driven approach is one of the earliest and most widely used approaches in stochastic occupant behavior models [28], [29]. The probability-sampling-driven approach employs probability distributions derived from input data to drive the simulation by sampling event characteristics (e.g., start time and duration) [30], [21]. The Markov chain approach is another widely adopted approach in stochastic simulation of residential building loads with varying occupant behavior. Markov chain approaches involving time can be categorized as time-homogenous or time-inhomogeneous, depending on whether the transition probability matrix varies across different time steps in the simulation [31]. Also,

Markov chains can be first-order, in which transition probabilities depend only on the current state (memoryless), or $n$th-order, in which transition probabilities depend on the current state and $n$ previous states [27].

These different approaches have been utilized to model various aspects of residential building usage, such as occupancy, HVAC, and lighting usage. Modeling occupancy (presence or absence) constitutes one of the first steps in stochastic modeling of occupant behaviors in the residential setting. Richardson et al. [32] established a high-resolution building occupancy model based on Markov chains and UK time-use survey data. The differences between weekdays and weekends were accounted for in the simulation. To improve the limited memory in the Markov chain simulation, Ramírez-Mendiola et al. [33] extended the first-order Markov chain into a variable-order Markov chain to model residential activities, including occupancy. Aerts et al. [34] developed a probabilistic approach to model occupancy sequences using Belgian time-use survey data. In addition to modeling the sensible and latent heat gain from occupants themselves, simulating realistic behaviors of occupant-driven appliance and hot water usage is also important, considering the large portion of household energy consumption attributed to appliances, lighting, and miscellaneous plug load usage [35]. As examples, Wilke et al. [36] derived probabilities of occurrence and duration of occupant activities and calibrated the simulation against French time-use survey data. Armstrong et al. [37] combined an occupant behavior simulation with appliance characteristics (power level, cycle features) to simulate the occupant-driven electricity load in Canada. As for lighting usage, Richardson et al. [38] simulated dwelling lighting usage and correlated it with occupancy and ambient lighting. Using a similar approach, McKenna and Keane [39] stochastically simulated lighting usage along with residential building loads in Ireland. For HVAC usage, Ren et al. [40] established stochastic action-based models with both environmental factors (indoor temperature) and events (presence/absence, sleep/awake) as triggers. Fabi et al. [41] also developed a logistics regression model to infer the thermostat set point for heating.

One major limitation of past behavior modeling efforts is that they have typically relied solely on time-use survey data, which limits modeled end uses to those found in the time-use data. For example, hot water usage of residents and its correlation with household activities are usually neglected. Previous efforts that focused solely on hot water usage include Hendron et al., who developed a tool for generating realistic residential hot water event schedules [42] as well a set of standardized hot water event schedules [43]. These have been widely used for simulation of solar water heaters, heat pump water heaters, and other water heating technologies (e.g., [44], [45]). Although these stochastic schedules have been implemented in software for modeling individual homes and multifamily buildings [46] as well as urban/national building stocks, they have not yet been applied at scale with sufficient heterogeneity to avoid artificially coincident peaks. Kruis et al. [47] took a different approach by using measured household water draw data and assembling sampled day profiles into annual profiles, but their use of 65 annual profiles does not achieve heterogeneity sufficient for building stock modeling. These prior applications are also limited in that they (1) do not align hot water usage patterns with other aspects of occupant behavior, and (2) do not achieve realistic consistency in behavior from day to day (each day is essentially a new household). Considering the significance of water heating (which comprises 14% of U.S. residential energy use) [48], it is important to model the hot water usage of different household activities (e.g., showering, clothes washing, dishwashing).

## 1.4 Contributions

In this paper, we propose an approach that combines timeseries clustering of U.S. time-use survey data with stochastic occupant behavior simulation. The stochastic occupant behavior simulation is integrated with a bottom-up physics-simulation model of the U.S. housing stock. After comparing the probability-sampling-driven approach and Markov chain approach for occupant behavior modeling, we recognize that the Markov chain approach, combined with duration sampling, achieves the best performance in realistically simulating resident behaviors. These simulated activities include all major end-use activities relevant to both electricity and hot water consumption. The behavior modeling is further integrated into ResStock™, a versatile building energy simulation platform developed to support different scales of residential building load analysis, ranging from county level to national level [49]. This work aims to establish a realistic and reliable stochastic model of residential building loads.

The remainder of the paper is organized into sections that (1) discuss the proposed approach in detail, (2) describe integration into ResStock, and (3) present discussions and conclusions.

# 2. Approach Overview

## 2.1 Modeling Framework

The proposed approach relies on the concept of agent-based modeling to stochastically simulate occupant behavior and associated household energy consumption. Each occupant is modeled as an agent, and each household is composed of one or more agents. First, we employ clustering on ATUS data to identify existing typical behavior patterns (Section 3.2). Next, we derive the Markov chain transition probability matrix for each cluster, along with a probability density function of activity durations for each cluster (Section 3.3). In the model application stage, we probabilistically assign one weekday and one weekend behavior pattern cluster to each occupant in a household. We then perform a Markov chain simulation for each occupant for each day of the year, along with sampling durations for each event. Finally, we aggregate these simulated behavior schedules to the household level and integrate them into ResStock, where they serve as inputs for stochastic residential building load profile modeling. EnergyPlus™ is used as the underlying building simulation engine. Figure 2 shows the general workflow of stochastic residential building load simulation in the training and application phase. We anticipate that future work will associate the behavior clusters with demographic variables such as age or employment status so that the occupant modeling can be localized for different geographic regions and perhaps correlated with housing types (such as single-family vs. multifamily housing units).

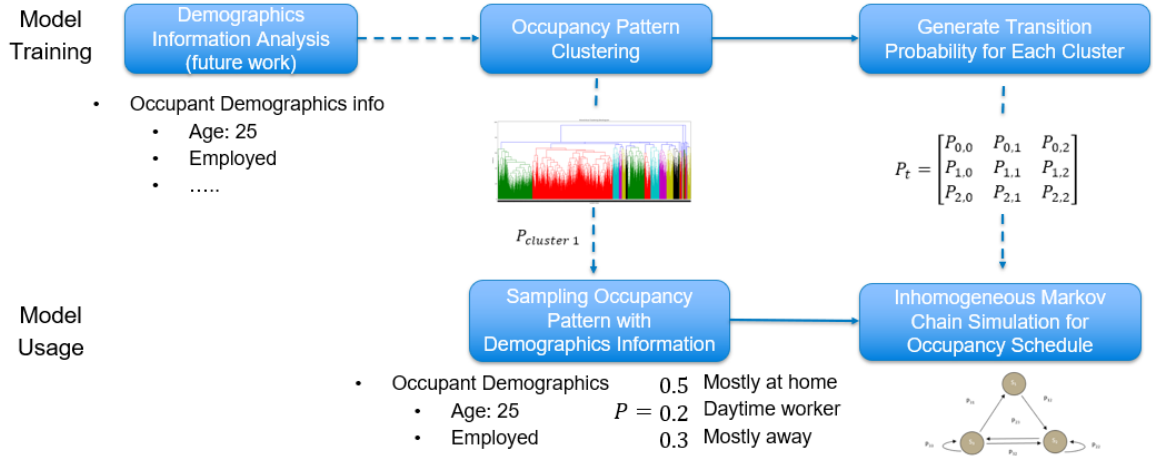

Figure 2. Training and application of stochastic occupant behavior modeling

## 2.2 Data Support – American Time Use Survey and Domestic Hot Water Event Generator

High-quality data are important for ensuring the accuracy and validity of stochastic models of occupant behavior. As one of the most detailed sources of recorded daily occupant behaviors, time-use survey data have been used by numerous studies to stochastically model occupant behaviors [50], [51]. In this research, we data from the ATUS [52]. The ATUS is conducted annually by the U.S. Bureau of Labor Statistics and collects data at minute-level resolution. Specifically, we use 5 years of data (from 2013 to 2017) with approximately 55,000 data records in total, covering a wide spectrum of occupant demographics, including age, employment status, and gender. The activity reporting time starts at 4:00 am and goes through to the same time the next day. The ATUS was collected with defined weights for each respondent so that the data are representative for the whole U.S. population. Hence, the ATUS was selected as the foundation for our stochastic occupant behavior modeling.

Although ATUS data are comprehensive with regards to occupant behavior recording, the ATUS only collects data on human activities and not on direct energy or water usage. For example, time recorded as "laundry" could include loading a clothes washer or folding laundry; it does not directly correspond to the duration of a clothes washer run cycle. Similarly, time recorded as spent on "personal hygiene" does not directly correspond with the duration of a shower hot water use event. Thus, we supplemented ATUS with other datasets that do provide this information. We derived distributions of clothes washer, clothes dryer, dishwasher, and electric range/stove/oven power levels and durations from the Northwest Energy Efficiency Alliance's 2011 Residential Building Stock Assessment (RBSA) Metering Study, which submetered end-use energy consumption of appliances and other circuits with 15-minute resolution in around 100 households [53]. We used domestic hot water event characteristics—probabilities for event start times, durations, and flow rates—that were previously derived by Hendron et al. from two Aquacraft studies [54]. The larger Aquacraft study measured the water use by fixture (hot and cold together) in 1,200 homes, and the smaller 20-home study measured hot and cold water use separately [55] [56]. The fixtures measured in both studies include sinks, showers, baths, clothes washers, and dishwashers.

# 3. Detailed Approach
## 3.1. Clustering of Occupant Behavior Patterns

We performed a clustering analysis to identify typical occupant behavior patterns. We ignored activity categories that were not relevant to the end uses modeled in ResStock, labeling them as "home/active" or "away." We also aggregated the 1-minute resolution data into 15-minute intervals.[2] Each processed occupant data point has 96 dimensions, representing the individual occupant behavior of one day in 15-minute intervals. The clustering was done based on just three categories: (1) home and active, (2) home and asleep, and (3) absent from home. Because time-use data are categorical, traditional clustering algorithms that work on continuous variables are not as applicable. Thus, we used the *k*-modes algorithm to recover and identify the underlying behavior patterns. The *k*-modes algorithm is an extension of the *k*-means algorithm, which forms data clusters based on the number of matching dimensions of data points [57]. With the advantages of fast computation and efficient memory usage, *k*-modes is applicable to the large-scale, high-dimensional data in this study [58]. The steps involved in *k*-modes clustering are as follows:

1. Initialization – randomly initialize the *k* number of cluster centers
2. Cluster assignment – assign all data points to the corresponding cluster centers based on the measured distance between the data point and the center
3. Center recalculation – recalculate the cluster center location as the average of all data points that belong to that cluster
4. Repeat Steps 2 and 3 until the algorithm converges, i.e., the cluster centers no longer shift.

To measure the similarity between data points, we used the Manhattan distance, as shown in equation 1.

$$D_{i,j} = \sum_{k=1}^{d} |x_i^k - y_j^k| \qquad (1)$$

The similarity of data points $i$ and $j$ is determined based on the category match in all $d$ dimensions of the data. For evaluating clustering performance, we utilized the silhouette score, which is one of the most widely used metrics in clustering evaluation. The silhouette score formula is presented in equation 2, in which $a$ is the mean distance between one sample and all other samples within the cluster, and $b$ is the mean distance between one sample and all samples in the next nearest cluster.

$$silhouette\_score = \frac{b - a}{\max(a, b)} \qquad (2)$$

Hence, the silhouette score measures the clustering performance by quantifying the average intra-cluster and inter-cluster distance. A larger silhouette score indicates more coherent intra-cluster data points with better separated clusters.

---

[2] Although ATUS allows 1-minute resolution in respondents' time-use diaries, many respondents round to 5-minute, 15-minute, 30-minute, or hourly resolution in their diaries.

Because the *k*-modes algorithm randomly initializes cluster centroids, we ran the algorithm 10 times and calculated the average silhouette score, with the number of clusters ranging from 3 to 10. Figure 3 presents the clustering performance evaluation results. The average silhouette scores for three clusters and four clusters are very similar to each other. The silhouette score experiences a drop at five clusters and consistently decreases as the number of clusters grows larger. Hence, the number of clusters was chosen as four to maintain a balance between clustering performance and diversity of occupant behaviors for later behavior simulation. Randomly chosen sets of 100 examples of daily occupant behavior in each of these four clusters are shown in Figure 4. We give the four clusters descriptive names, as follows: (1) mostly at home; (2) mostly at home, early riser; (3) daytime away, evenings home; and (4) daytime away, evenings away.

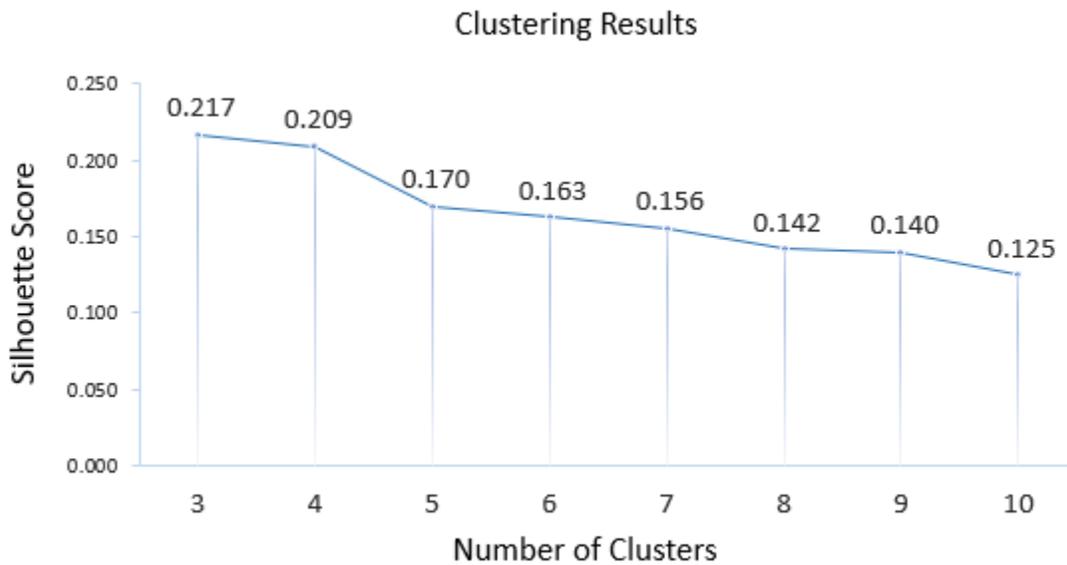

*Figure 3. Clustering performance*

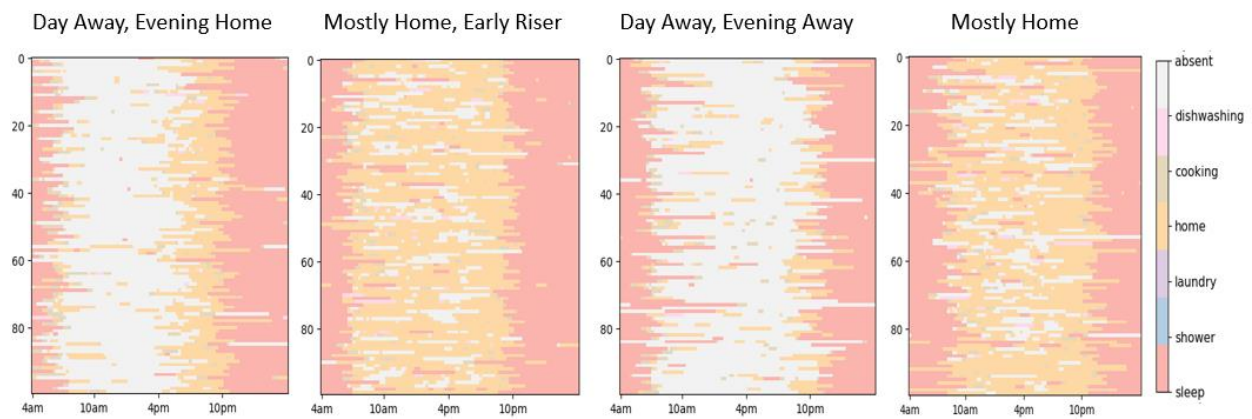

*Figure 4. The clustering analysis resulted in four weekday behavior clusters, illustrated here with 100 occupant-day instances. 36% of ATUS respondents are in cluster 1 (day away, evening home), 21% are in cluster 2 (mostly home, early riser), 21% are in cluster 3 (day away, evening away), and 22% are in cluster 4 (mostly home). Although only "absent," "sleep," and "home/active" were used for clustering, other activity types are shown for context.*

## 3.2. Behavior Modeling

### 3.2.1 Major Existing Approaches in Behavior Modeling

As summarized in the literature review, the two most common approaches to stochastic behavior modeling are probability sampling and Markov chains.

*Probabilistic Sampling*

In the typical probabilistic sampling approach, researchers sample the number, start time, and duration of events from probability density functions [26], [34]. If the event is "eating a meal," then this approach is equivalent to asking, "How many meals should I eat today?" and then for each meal, asking, "What time should I eat, and how long should it last?"

A second commonly used probabilistic sampling approach is to sample the activity likelihood at each time step from a probability profile, which depicts the activity-on probability across a day [21], [37]. This is equivalent to asking, "Am I eating at 6:30 pm?" and then, "Am I eating at 6:45 pm?" at the next time step, and so on. One limitation of this approach is that it does not provide control over the number of events that occur. Despite the ease of implementation, in both types of probabilistic sampling, the simulated occupant end-use activity could become split up into multiple events.

*Markov Chain*

Another widely used behavior modeling approach is the Markov chain. It models a sequence of consecutive events based on probabilistic rules to determine the state transition [27]. The probabilistic rules, depicted as transition probability matrices and the core of Markov chains, define the probability of transitioning from one state to another from time step $t$ to $t+1$. This is demonstrated in equation 3. As an example, three states are defined in the transition probability matrix $P_{t,t+1}$.

$$P_{t,t+1} = \begin{bmatrix} P_{11} & P_{12} & P_{13} \\ P_{21} & P_{22} & P_{23} \\ P_{31} & P_{32} & P_{33} \end{bmatrix} \text{ with } P_{ij} = P(x_t = s_j | x_{t+1} = s_i) \quad (3)$$

Each element in the matrix dictates the probability of transiting from one state to another. Hence, in the Markov chain simulation, the future state is dependent on previous states. The activity dependency could range from one to several time steps, representing different orders of Markov chain (i.e., first order means the activity in the current time step is dependent only on the activity in the previous time step). In addition, Markov chains can be classified as time-homogeneous or time-inhomogeneous. The transition probability matrix varies across different time steps of the simulation for time-inhomogeneous Markov chains, but stays constant for time-homogeneous Markov chains. A first-order time-*homogeneous* Markov chain can be thought of as answering the question, "If I am currently cooking, what is the probability that I will eat next?", whereas a first-order time-*inhomogeneous* Markov chain can be thought of as answering the question, "If I am currently cooking and it is 6:15 pm, what is the probability that I will be eating at 6:30 pm?"

In this study, we employed first-order, time-inhomogeneous Markov chains, considering the experience of past building occupant modeling efforts, which primarily use first-order Markov chains. We performed the simulation with 15-minute time steps to balance simulation resolution and training data size requirements.

### 3.2.2 Approach Evaluation for Behavior Modeling

To ensure the validity and accuracy of our behavior modeling, we evaluated three approaches: one pure Markov chain approach, and two hybrid approaches that combine Markov chains with elements of probabilistic event sampling. Figure 5 shows a diagram comparing the three approaches evaluated. The simulation performance was evaluated by comparing distributions of activity start time, activity duration, number of occurrences per day, and overall activity probability profile (for 10,000 simulated occupant days) to the equivalent distributions from reported ATUS data (N=27,408 for weekdays and 27,190 for weekends). In the next section, we will introduce the details and comparison of each evaluated approach. Then, we will present the validation of the selected approach.

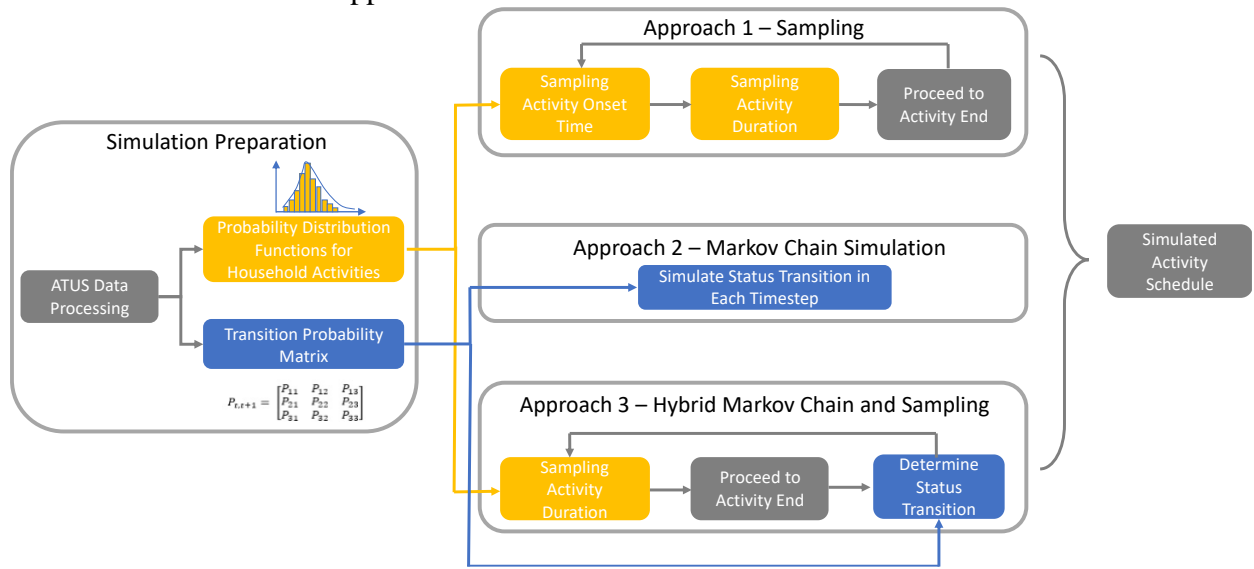

*Figure 5. Diagram comparing the three approaches evaluated*

#### *Approach 1 – Markov Chain with Probabilistic Sampling*

The first approach we investigated was to combine a Markov chain for occupancy status (active/away/sleep) with probabilistic sampling of onset time, duration, and number of occurrences per day for specific activities. We processed the ATUS data to derive (1) the Markov chain transition probability for occupancy status (active/away/sleep) simulation, and (2) the probability density functions of activity start time, duration, and number of occurrences per day.

For each occupant, we used a Markov chain to simulate the occupancy status. Based on the simulated occupancy status, we identified the available periods (i.e., when the occupant is active and at home) for the occurrence of different activities (e.g., laundry, dishwashing, cooking) and sampled the number of activity occurrences, onset times, and durations in the available periods. This is to ensure that simulated occupant energy use behavior happens only when the occupant is at home and active, while reducing Markov chain training complexity. For this hybrid approach, we found a good match between simulated and actual activity durations and number of activity occurrences, but large differences in activity onset times and daily probability profiles. Figure 6 shows an example of these discrepancies for the simulation of laundry.

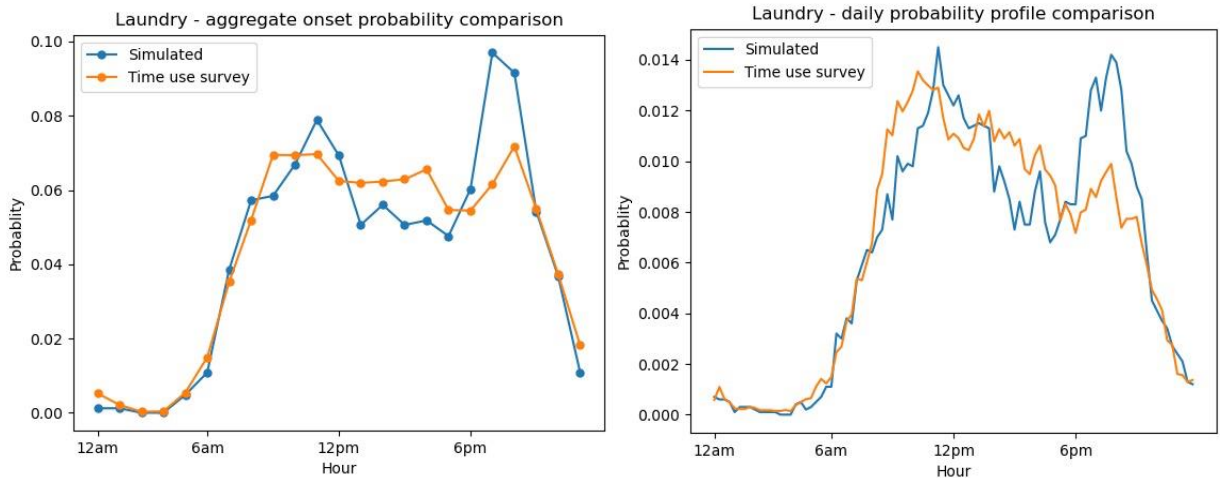

*Figure 6. Although Approach 1 results in good agreement between simulated and actual activity durations and number of activity occurrences, it results in large differences in activity onset times and daily "on time" profiles*

### Approach 2 – Pure Markov Chain

The second approach we evaluated was a pure Markov chain simulation for all occupant activities with energy and hot water usage implications, including sleep, being away from home, laundry, dishwashing, personal hygiene, cooking, and other home/active activities (e.g., reading, eating, and so on). Comparing the simulated behavior schedules with the actual ATUS data, we found that in general, the full activity Markov chain could reproduce the occupant behavior realistically. However, for certain activities such as laundry, the simulated activity durations had large discrepancies compared to the ATUS data, as demonstrated in Figure 7. The concentration of laundry activity durations in the 1.0-, 1.5-, and 2.0-hour values could be due to a rounding bias in ATUS respondent reporting, so this discrepancy is not necessarily a negative result. Nevertheless, we strove to develop an approach that could match the duration profile more precisely.

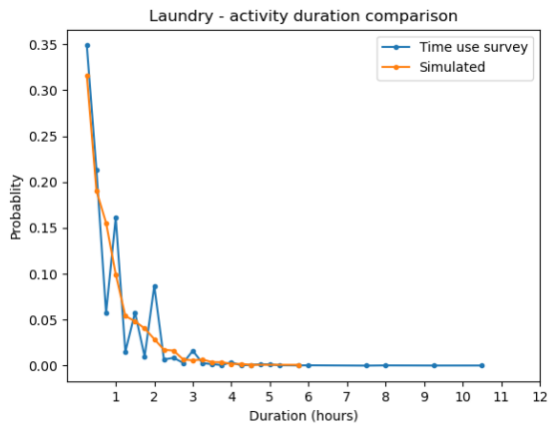

*Figure 7. Approach 2 struggled to reproduce the discrete profile of laundry activity duration*

*Approach 3 – Markov Chain with Duration Sampling*

Informed by the observed discrepancies in Approaches 1 and 2, we developed a new approach that takes the best features of each: using the full activity Markov chain for energy-related activity onset times (as in Approach 2) and using probabilistic sampling for the duration of activities (as in Approach 1).

An example of this approach for the simulation of one occupant-day is:

1. Day starts at 4:00 AM. Use a Markov chain to determine initial activity at 4:00 a.m. → **sleep**
2. Use the Markov chain to determine next activity at 4:15 a.m. → again, **sleep**
3. Keep using the Markov chain to get next activity → **sleep**, until 6:45 a.m.
4. Use the Markov chain at 6:45 a.m. to determine next activity → **personal hygiene**
5. Sample **personal hygiene** duration from a probability distribution → 60 minutes
6. Probabilistically chose if the **personal hygiene** is a shower or bath → **shower**
7. Sample **shower** duration from a probability distribution → 9 minutes
8. Randomly select **shower** start time from within a 60-minute **hygiene** period with 1-minute resolution → **shower** from 6:47 a.m. to 6:56 a.m.
9. Use the Markov chain, which resumes at end of **personal hygiene** period (7:45 a.m.), to determine next activity.

As demonstrated in the next section, Approach 3 performs better than Approaches 1 and 2. Additionally, because some ATUS activities (e.g., personal hygiene and laundry) do not directly correspond to energy or hot water use events, we already need to sample event duration for those activities (using non-ATUS datasets). Therefore, selecting Approach 3 means that durations will be sampled for all activities in a consistent manner (some using ATUS-derived durations and some using non-ATUS-derived durations).

*Validation of the Selected Approach*

The simulated and actual behavior schedules were compared for all relevant activities, including cooking, dishwashing, laundry, and personal hygiene. We compared the activity onset time probability, activity duration, number of occurrences, and the resulting daily probability profiles. The results of these evaluations are shown in Figure 8.

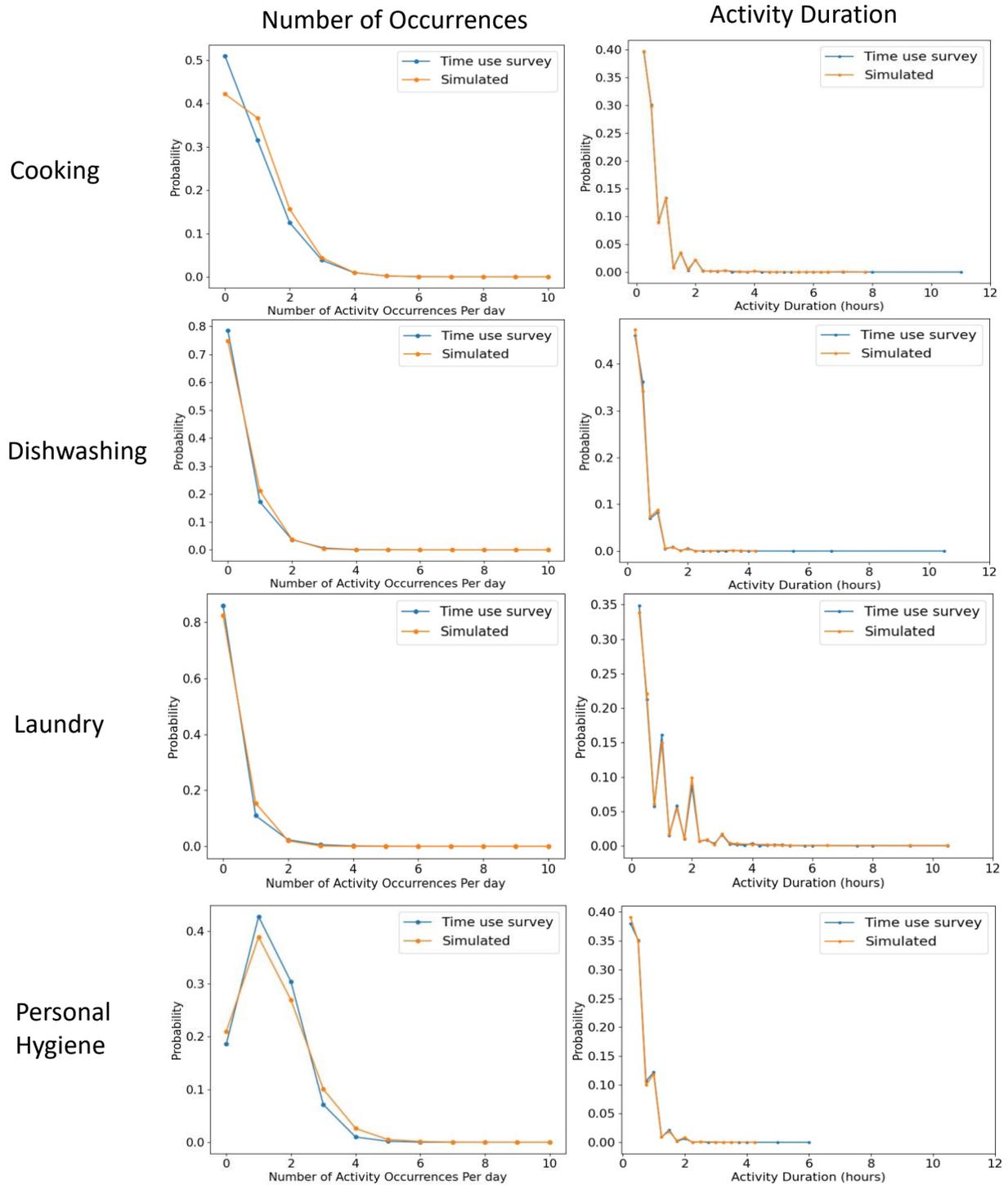

*Figure 8. Validation of Approach 3, showing agreement in number of occurrences and duration for four energy-related activities*

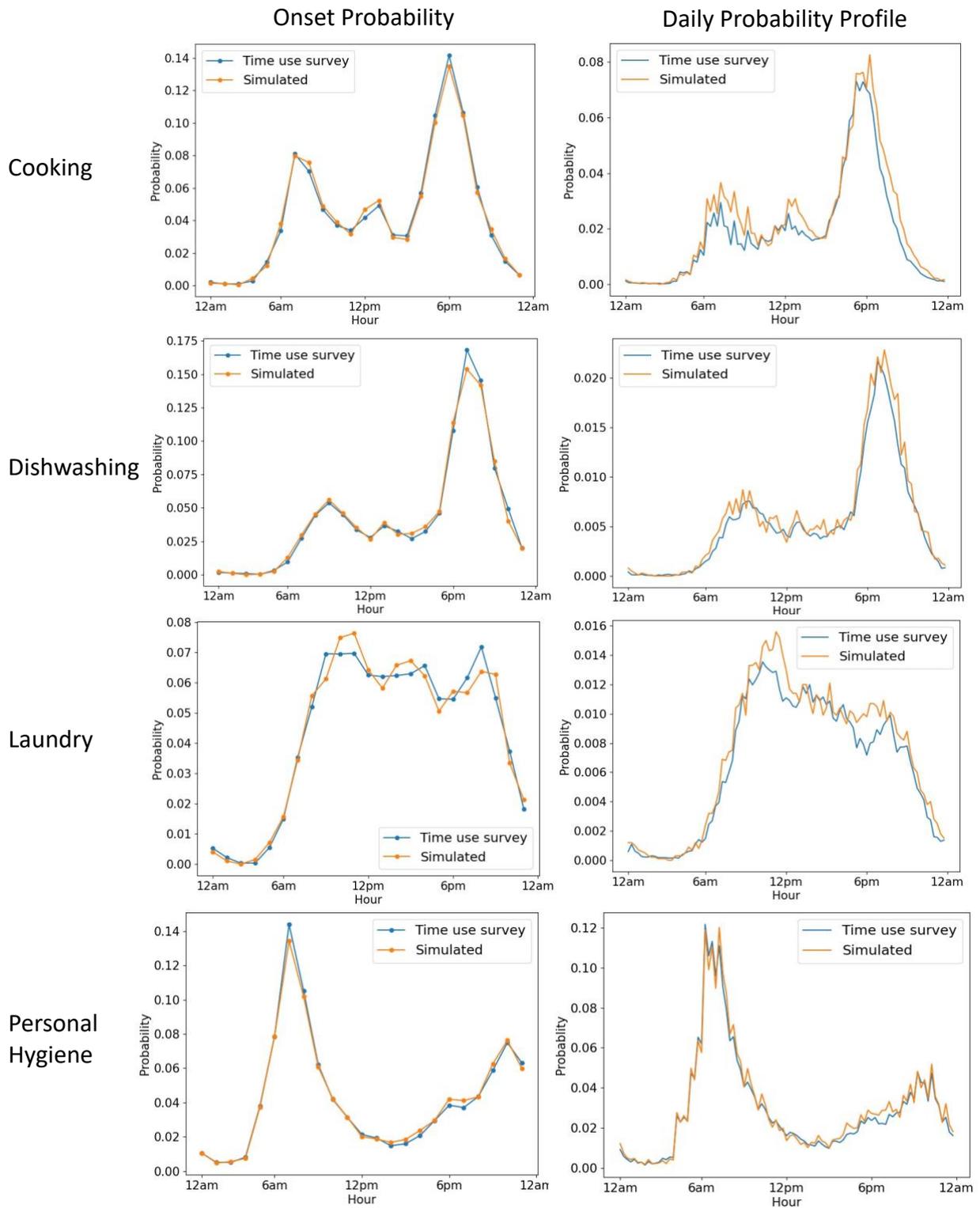

*Figure 9. Validation of Approach 3, showing agreement in activity onset time probability and the resulting daily probability profile for four energy-related activities*

An accurate simulation of occupants' presence (sleep, home and active, away) is also significant, because it is fundamental to energy consumption of households. To validate the simulated occupancy status, we compared the daily probability profile, status durations, and number of status changes per occupant per day between the ATUS data and the simulated schedules, as demonstrated in Figure 10. The occupant status durations and number of status changes were compared using cumulative distribution functions. We observed that the simulated and real data matched well in all comparisons.

Compared to Approach 1 and 2, which have visible discrepancies in daily probability profiles or activity durations, Approach 3 demonstrates more accurate simulation of all aspects of occupant activities. Thus, we selected Approach 3 as the behavior modeling approach in this study.

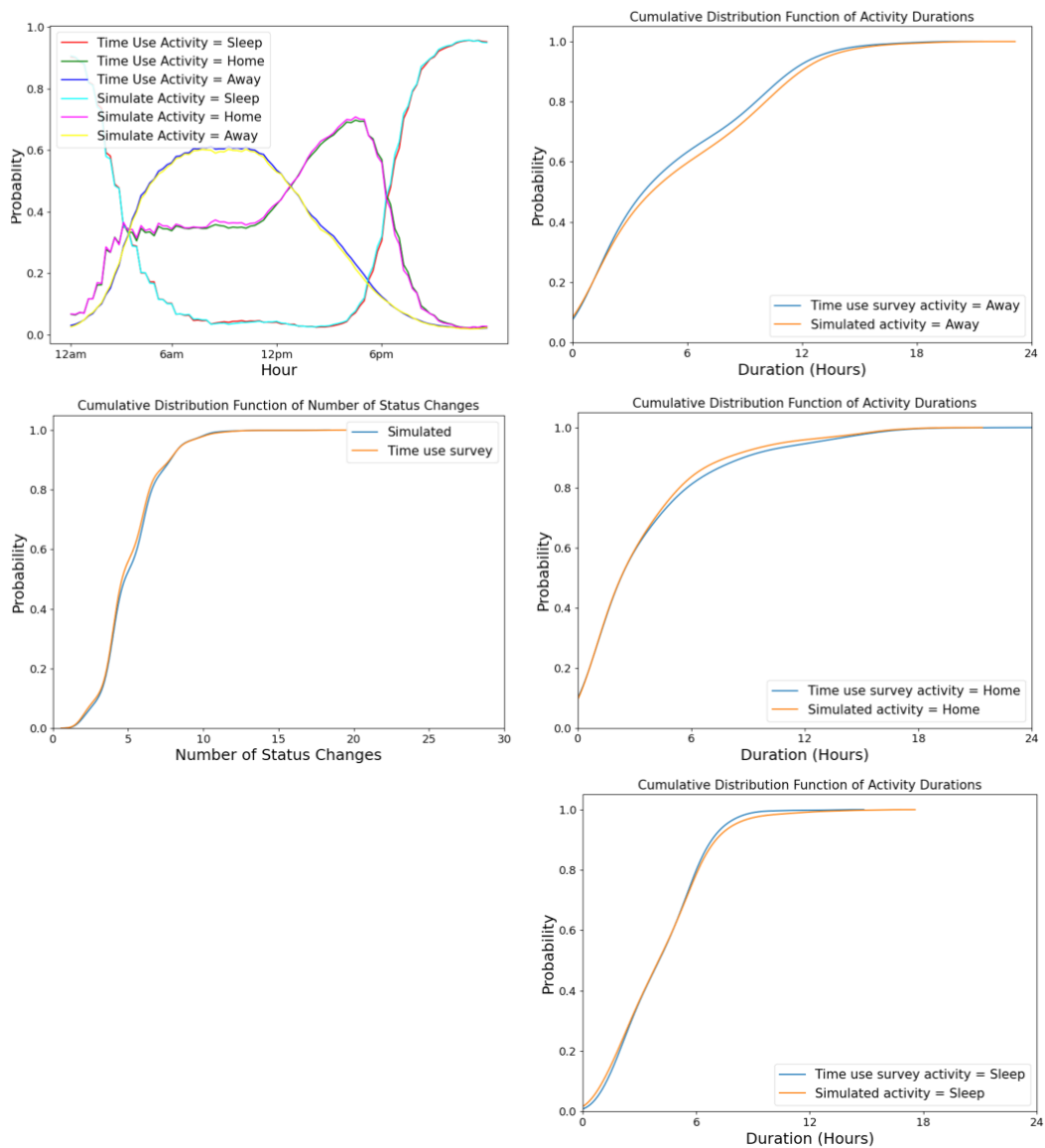

*Figure 10. Validation of simulated occupancy status (sleep, home and active, away) using Approach 3, showing agreement in overall status probabilities, number of status changes, away duration, home/active duration, and sleep duration*

### 3.2.3 Generating Schedules From Simulated Behavior

This section describes how simulated behavior for individual occupants is aggregated into activity schedules for whole households. The complete procedure for household activity schedule aggregation is shown in Figure 11.

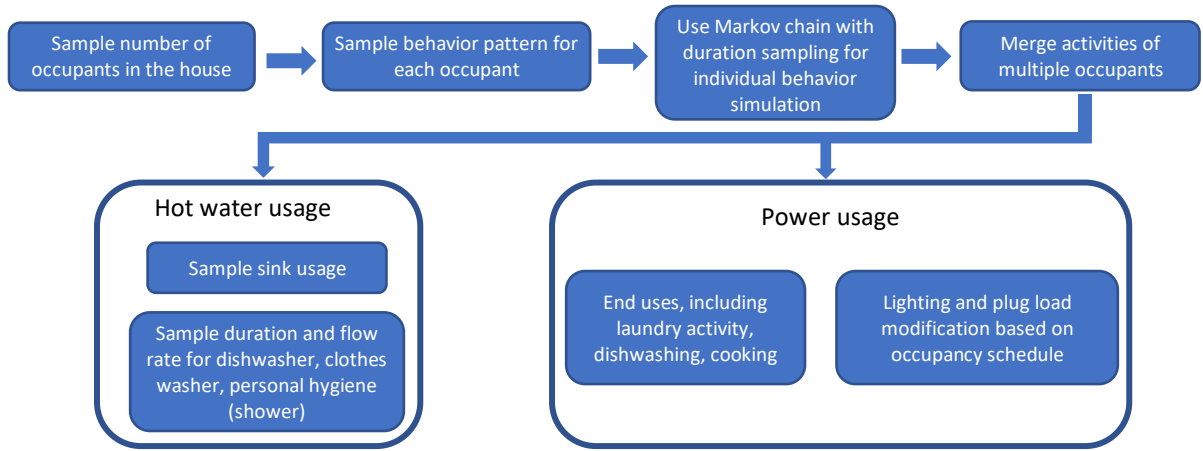

*Figure 11. Flowchart for household level activity simulation/aggregation*

Table 1 summarizes how ATUS activities are mapped to various EnergyPlus building simulation schedules, including both power and water draw schedules. It also shows whether the schedule generation is done at the occupant level or at the household level, whether overlapping individual occupant events are combined into a single household event, and which data sources are used to sample event onset time, duration, and magnitude of power draw or water draw. Most schedules needed for building energy modeling of a household can either be classified as power draw schedules or hot water draw schedules. For household events such as laundry or dishwashing, the duration of power draw is not the same as the duration of water draw. Hot water usage typically causes electricity or gas use of a tank water heater at a later time (tankless water heater use is more closely aligned to the water draws). Thus, it is important to separate hot water draw profiles from power draw profiles.

*Table 1. Summary of relevant ATUS activities and their associated EnergyPlus schedules, whether the schedule generation is done for one occupant or for the entire household, whether overlapping individual occupant events are combined into a single household event, and which data sources are used for event onset time, duration, and magnitude of power draw or water draw*

| Relevant ATUS Activity | EnergyPlus Building Simulation Schedule | Type | Combine Overlapping Events Into Single Event? | Data Sources | | |
|---|---|---|---|---|---|---|
| | | | | Start Time | Duration | Magnitude (Power, Flow) |
| Away | None | Occupant | No | ATUS | ATUS | None |
| Sleeping | None | Occupant | No | ATUS | ATUS | None |
| Any except "away" | Heat gain - occupants | Occupant | No | ATUS | ATUS | ASHRAE |
| Personal hygiene | Water - showers | Occupant | No | ATUS | DHWESG | DHWESG |
| Personal hygiene | Water - baths | Occupant | No | ATUS | DHWESG | DHWESG |
| Dishwashing | Water - dishwasher | Occupant | Yes | ATUS | DHWESG | DHWESG |

| | | | | | | | |
|---|---|---|---|---|---|---|---|
| Dishwashing | Power - dishwasher | Occupant | Yes | ATUS | RBSAM | RBSAM |
| Laundry | Water - clothes washer | Occupant | Yes | ATUS | DHWESG | DHWESG |
| Laundry | Power - clothes washer | Occupant | Yes | ATUS | RBSAM | RBSAM |
| Laundry | Power - clothes dryer | Occupant | Yes | ATUS | RBSAM | RBSAM |
| Cooking | Power - cooking range | Occupant | Yes | ATUS | ATUS | RBSAM |
| Not away or sleeping | Water – sinks | Household | N/A | DHWESG | DHWESG | DHWESG |
| Not away or sleeping | Power - misc. electric loads | Household | N/A | Prorate avg. schedule based on occupancy | | |
| Not away or sleeping | Power - lighting | Household | N/A | Prorate avg. schedule based on occupancy | | |
| Not away or sleeping | Power - ceiling fans | Household | N/A | Prorate avg. schedule based on occupancy | | |
| N/A | Heating thermostat | Household | N/A | RECS (not yet connected to occupancy) | | |
| N/A | Cooling thermostat | Household | N/A | RECS (not yet connected to occupancy) | | |

ATUS = American Time Use Survey
ASHRAE = ASHRAE Handbook–Fundamentals (2009)
DHWESG = NREL Domestic Hot Water Event Schedule Generator (based on Aquacraft data)
RBSAM = Northwest Energy Efficiency Alliance's 2011 Residential Building Stock Assessment (RBSA) Metering Study
RECS = U.S. Energy Information Administration's 2009 Residential Energy Consumption Survey (RECS)

For each run of ResStock, we sample the number of occupants in a household based on the building type (e.g., single-family, multifamily), number of bedrooms, and public use microdata area, which are 2,378 aggregations of U.S. census tracts and counties that contain at least 100,000 people. These probabilities are derived from the 2017 5-year American Community Survey Public Use Microdata Sample (PUMS) [59]. Next, we sample the corresponding number of occupants and associated living patterns for each occupant based on identified clusters of behavior patterns (as presented in Figure 4). Once the occupant behavior pattern is determined, a Markov chain simulation with duration sampling is performed for each occupant for each day of the year, using one transition probability matrix for weekdays and another for weekend days. These individual activities of household members were merged into household activity schedules. When two or more occupants have overlapping activities that would use the same household appliances (cooking, laundry, and dishwasher), the overlapping schedules are combined to avoid unrealistic power draws or water draws for a single appliance. Then, the household appliance power draw durations are sampled from probability distributions derived from the sources listed in Table 1.

### 3.2.4 Lighting, Plug Loads, and Ceiling Fans

The schedules for lighting, plug loads, and ceiling fans are generated in a two-step process. We start with a reference schedule that is derived from measured circuit-level data, and then we modulate the reference schedule based on the household occupancy, normalized as a fraction of maximum occupancy using the following equation:

$$x_{modulated,i} = x_{daily\ min.} + (x_{reference,i} - x_{daily\ min.}) \times occupancy_{norm.,i}$$

where

$x_{modulated,i}$ is the modulated schedule value during time step $i$
$x_{daily\ min.}$ is the minimum of the reference schedule for that day

$x_{reference,i}$ is the reference schedule value during time step $i$

$occupancy_{norm.,i}$ is the normalized occupancy fraction during time step $i$.

The implication is that, when the occupancy is full (1.0), the modulated schedule will have the same value as the reference schedule. When occupancy is zero, the modulated schedule is set to the minimum of the reference schedule., and when occupancy is partial, it will have a value between the daily minimum and the original reference value. The lighting and plug load schedules are normalized before they are used, and thus, this modulation changes only the timing (and not the annual consumption) of the lighting and plug loads, which are separate inputs. Figure 12 shows an example of the modulated lighting schedule for one day, created from the reference schedule using the simulated household occupancy fraction.

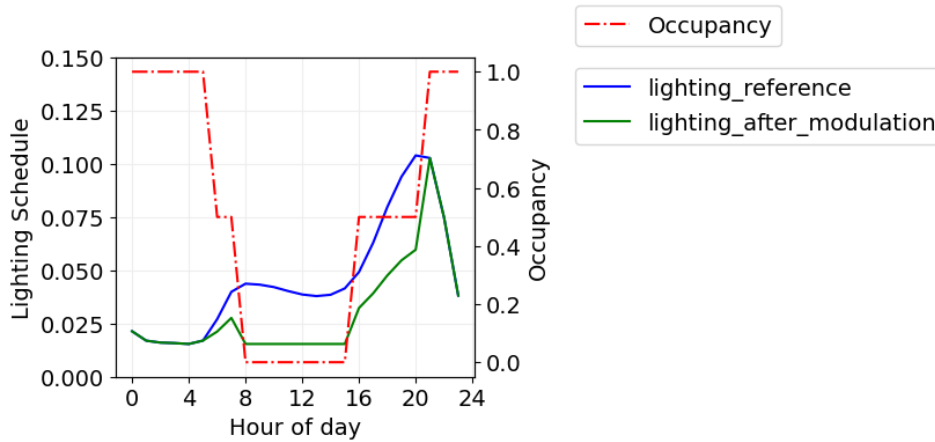

Figure 12. Example of lighting schedule modulated by the simulated household occupancy fraction

### 3.2.5 Hot Water Usage

For each hot water related activity occurrence in the combined household schedule (personal hygiene, dishwashing, laundry), the hot water draw duration and flow rate are sampled from probability distributions previously derived by Hendron et al. [52]. Personal hygiene events are probabilistically assigned to be either showers (92.1%) or baths (7.9%).

Separately, because the sink usage is not explicitly reported in the ATUS, the sampling of sink activity completely relies on Hendron et al.'s [52] probability distributions for sink use onset, draw clusters, flow rate, and draw durations. To ensure the reality of behavior simulation, the sink usage is only allowed to be activated when at least one occupant is in the active status at home. In this way, we generate the household activity schedules to serve as inputs for residential building load simulation by ResStock, which will be described in detail in Section 4.1.

## 4. Integration with ResStock for Residential Building Load Modeling

### 4.1 ResStock Introduction

ResStock is a U.S. Department of Energy model of the U.S. residential building stock, developed and maintained by the National Renewable Energy Laboratory (NREL) since 2014. ResStock was developed to represent the energy use and energy saving potential of residential building stocks at national, regional, and local scales with a high degree of granularity. Compared to other building stock models that use a handful of archetype or prototype models, ResStock typically uses many representative models—around 550,000 for the contiguous United States—to represent the residential building stock with high fidelity. Unlike many urban building energy modeling approaches, ResStock does not attempt to generate a physics model for every building, but rather uses a relatively large number of statistically sampled models to represent the building stock with a realistic diversity of building characteristics. ResStock is classified as a Q4 physics-simulation type of building stock energy model [15].

The ResStock methodology is depicted in Figure *13* and can be summarized as follows. (1) The building stock is characterized by a set of conditional probability distributions queried from more than a dozen public and private data sources. (2) With each ResStock run, representative housing units are sampled from the parameter space using deterministic quota sampling with probabilistic combination of noncorrelated parameters. (3) The housing units (both the baseline stock and the stock with application of energy efficiency or demand flexibility upgrades) are simulated, typically using the OpenStudio® software development kit [60] and associated residential modeling workflows for model construction and articulation, and the EnergyPlus® simulation engine for sub-hourly building physics simulation [61]. The large number of simulations requires running on high-performance or cloud computing resources. (4) Calibration/validation for ResStock was initially completed for annual single-family results in 2015; timeseries validation is ongoing in a 2019–2021 effort [12]. (5) Timeseries results from the potentially millions of baseline and upgrade scenarios can be extremely large and difficult to work with, so NREL has developed a stack of technologies to facilitate processing, aggregation, and analysis.

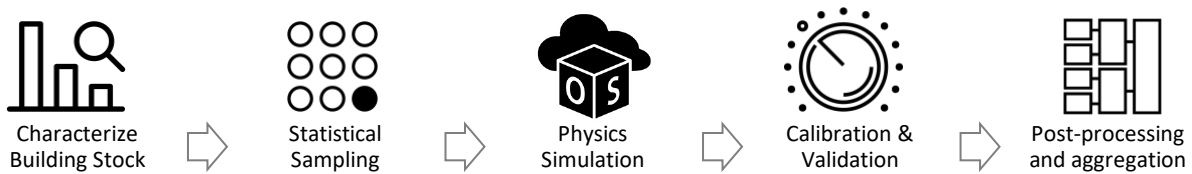

*Figure* 13. *ResStock methodology*

### 4.2 Occupancy Behavior Modeling Integration

The ResStock building energy model construction workflow is comprised of a series of OpenStudio measures, which are Ruby scripts for OpenStudio model articulation. The new stochastic modeling capabilities are implemented in a new measure called ResidentialScheduleGenerator. As shown in Figure 14, this measure is called early in the ResStock measure workflow. It takes as inputs the number of occupants, vacation start time, and vacation end time arguments selected via the ResStock sampling process. It uses two pre-trained Markov chain transition probability matrices (weekday and weekend) for each of the four occupant behavior types identified from the clustering process. After the occupant behavior schedule is generated for each 15-minute interval for the year, the actual appliance schedule and hot water fixture schedule are generated by combining the behavior schedule with the appliance power probability distribution and hot water fixture water draw probability distributions (as documented in Table 1). The output of the measure is a comma-separated value (csv) schedule file containing the normalized schedules for occupancy level (the proportion of the total number of occupants present in the home at each time step), and appliance power and water draws by various appliances and fixtures. The residential building geometry measures for single-family and multifamily housing units apply an occupant heat gain schedule to the living space by reading the occupancy level schedule from the csv file. The electric appliance measures use the power schedules and the hot water fixture measures use the hot water draw schedules from the csv file to apply occupant behavior-based usage schedules to the EnergyPlus component models.

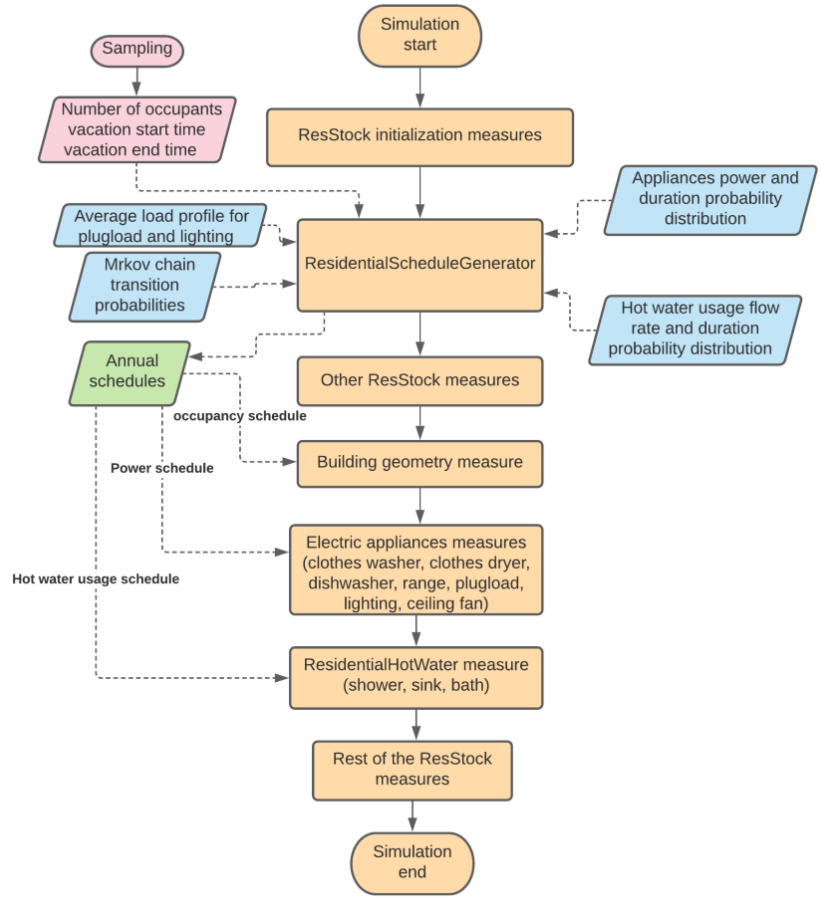

Figure 14. Flowchart showing how the occupancy-based schedule generation is implemented in ResStock

## 4.3 Simulated Residential Load Profile Validation

To validate our stochastic occupant behavior simulator, we compared ResStock simulation results for the Pacific Northwest region to measured data for six end uses from the Northwest Energy Efficiency Alliance's 2011 Residential Building Stock Assessment (RBSA) Metering Study [53]. The plots in Figure 15 show normalized average weekday and weekend profiles for ResStock results and RBSA Metering data. The RBSA Metering dataset is shown as a shaded region representing the 95% confidence interval of the mean ($mean \pm 1.96 \times standard\ error$). The number of samples for each end use ranged from 57 to 103 and are shown in the figure.

The shapes of the simulated profiles are generally similar to the shapes of the measured profiles, with some differences (e.g., dishwasher and water heating) that may be explained by a bias in the sample homes measured in RBSAM compared to the larger national sample in ATUS activity profiles. Some minor timing differences may be explained by the fact that ATUS activities (e.g., washing dishes and doing laundry) do not directly correspond to when energy is used for dishwashers, clothes washers, and clothes dryers.

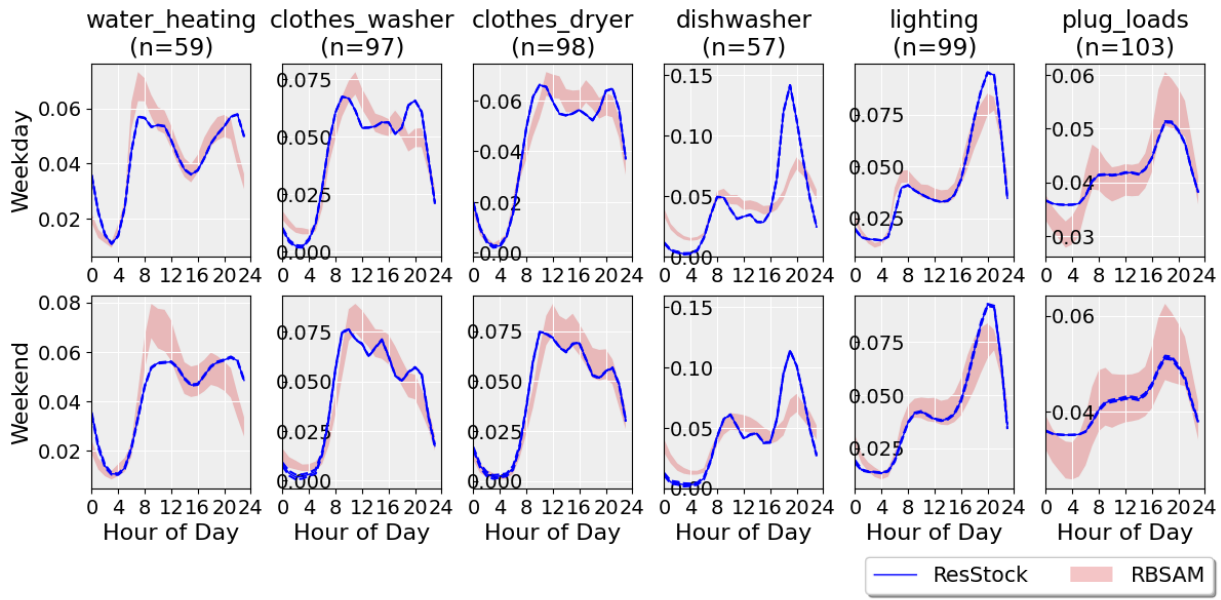

Figure 15. Comparison of normalized average weekday and weekend profiles for ResStock simulation results and RBSAM measured data, which is shown as a 95% confidence interval of the mean ($mean \pm 1.96 \times standard\ error$) for six end uses

## 5. Conclusion and Future Work

We developed a stochastic occupant behavior simulator and integrated it with a large-scale building stock simulation platform to realistically simulate the heterogeneity and stochasticity of residential building loads at scale. To ensure the diversity of occupant behavior in the simulation, we used clustering techniques to identify typical behavior patterns. Then, we evaluated and compared three approaches, ultimately selecting a hybrid approach that uses Markov chains for activity selection and probabilistic sampling for event durations. We integrated this stochastic behavior simulator as an OpenStudio measure for the ResStock building stock simulation framework, which allows to stochastically model the residential occupant-driven loads at scale. We verified the reliability of the occupant behavior simulator by comparing it against ATUS data. Finally, we validated the simulated residential building loads against circuit-level metering data from the RBSA Metering study.

Planned extensions of the model include incorporating demographic information into the workflow to account for correlations between demographic variables such as age or employment status and occupant behavior. In addition, we plan to integrate heating and cooling thermostat set point schedules with the simulated occupancy status to make set point behavior, like setbacks while occupants are sleeping or away from home, consistent with the other simulated occupant-driven energy use.

The ResStock-integrated stochastic occupant behavior simulator presented here represents a major advancement in residential occupant simulation. Many occupancy simulators use time-use survey data, but the simulator presented here is unique in that it combines data from multiple sources to model the effect of occupant behavior on electricity, gas, and hot water use consistently across a wide range of end uses, including clothes washing, clothes drying, cooking, dishwashing, sink use, showers, baths, lighting, and plug loads. Additionally, the simulator has been integrated with a national-scale building stock simulation framework and validated against circuit-level submetering electricity use data.

A forthcoming publicly available dataset of end-use load profiles for the U.S. building stock will incorporate the advancements presented here. The more realistic heterogeneity and stochasticity will improve accuracy for a wide range of analysis use cases, including utility rate design, ratepayer impact, non-wires alternatives, and electric vehicle hosting capacity, as well as solar photovoltaic hosting capacity, solar photovoltaic cost effectiveness, behind-the-meter battery/inverter sizing, microgrid design, and demand response and flexibility potential studies.

## Acknowledgements

This work was authored in part by the National Renewable Energy Laboratory, operated by Alliance for Sustainable Energy, LLC, for the U.S. Department of Energy (DOE) under Contract No. DE-AC36-08GO28308. Funding provided by U.S. Department of Energy Office of Energy Efficiency and Renewable Energy Building Technologies Office. The views expressed in the article do not necessarily represent the views of the DOE or the U.S. Government. The U.S. Government retains and the publisher, by accepting the article for publication, acknowledges that the U.S. Government retains a nonexclusive, paid-up, irrevocable, worldwide license to publish or reproduce the published form of this work, or allow others to do so, for U.S. Government purposes. A portion of this research was performed using computational resources sponsored by the Department of Energy's Office of Energy Efficiency and Renewable Energy and located at the National Renewable Energy Laboratory.

# CRediT author statement

Jianli Chen: Methodology, Software, Validation, Formal analysis, Writing - Original Draft

Rajendra Adhikari: Methodology, Software, Validation, Writing - Original Draft

Eric J.H. Wilson: Conceptualization, Methodology, Writing - Original Draft, Writing - Review & Editing, Supervision, Project administration, Funding acquisition

Joseph J. Robertson: Software

Anthony D. Fontanini: Conceptualization, Methodology, Supervision, Validation

Benjamin Polly: Conceptualization, Writing - Review & Editing, Supervision, Funding acquisition

Opeoluwa Olawale: Investigation, Writing - Review & Editing